\documentclass[12pt]{article}

\usepackage{scicite}

\usepackage{times}

\usepackage{graphicx}

\newenvironment{sciabstract}{%
\begin{quote} \bf}
{\end{quote}}

\newcounter{lastnote}
\newenvironment{scilastnote}{%
\setcounter{lastnote}{\value{enumiv}}%
\addtocounter{lastnote}{+1}%
\begin{list}%
{\arabic{lastnote}.}
{\setlength{\leftmargin}{.22in}}
{\setlength{\labelsep}{.5em}}}
{\end{list}}

\title{Negative differential conductivity in an interacting quantum gas}

\author
{Ralf Labouvie,$^{1}$ Bodhaditya Santra,$^{1}$ Simon Heun,$^{1}$ Sandro Wimberger,$^{2,3,4}$ \\Herwig Ott$^{1\ast}$\\
\\
\normalsize{$^{1}$Research Center OPTIMAS, Technische Universit\"at Kaiserslautern,}\\
\normalsize{67663 Kaiserslautern, Germany}\\
\normalsize{$^{2}$Institut f\"ur Theoretische Physik, Universit\"at Heidelberg,}\\
\normalsize{69120 Heidelberg, Germany}\\
\normalsize{$^{3}$Dipartimento di Fisica e Scienze della Terra, Universit\`{a} di Parma, Via G. P. Usberti 7/a,}\\
\normalsize{43124 Parma, Italy}\\
\normalsize{$^{4}$INFN, Sezione di Milano Bicocca, Gruppo Collegato di Parma, Italy}\\
\\
\normalsize{$^\ast$To whom correspondence should be addressed; E-mail:  ott@physik.uni-kl.de.}
}

\date{}

\begin{document} 


\baselineskip24pt


\maketitle


\begin{sciabstract}
Negative differential conductivity (NDC) is a widely exploited effect in modern electronic components. Here, a proof-of-principle is given for the observation of NDC in a quantum transport device for neutral atoms employing a multi-mode tunneling junction. The transport of the many-body quantum system is governed by the interplay between the tunnel coupling, the interaction energy and the thermodynamics of intrinsic collisions, which turn the coherent coupling into a hopping process. The resulting current voltage characteristics exhibit NDC, for which we identify a new microscopic physical mechanism. Our study opens new ways for the future implementation and control of complex neutral atom quantum circuits.
\end{sciabstract}


\section*{Main Text}
Transport of mass and charge is of fundamental importance in many physical, chemical and biological systems. On a microscopic level, many transport phenomena involve tunneling through a barrier, such as Josephson junctions \cite{Josephson1962} and squids \cite{Jaklevic1964}, tunneling diodes \cite{Esaki1958}, the tunneling to hopping transition in molecular wires \cite{Choi2008,Nozaki2012}, electron transport in DNA \cite{Winkler2005} or proton tunneling in enzyme reactions \cite{Cha1989}. Tunneling transport in solids typically occurs when an electron reservoir with chemical potential (or Fermi energy) $\mu_1$ is separated by a tunneling barrier or a quantum well from a second reservoir with chemical potential (or Fermi energy) $\mu_2$. This model and extensions of it have been introduced by Esaki and Tsu to calculate the conductance of resonant tunneling diodes and has culminated in the discovery of negative differential conductivity (NDC) \cite{ET1970}, a phenomenon widely exploited in electronic devices. It has also been observed, e.g., in single molecule junctions \cite{Perrin2014}, carbon nanotubes \cite{Pop2005}, and graphene transistors \cite{Britnell2013}. Ultracold quantum gases are predestined to simulate complex condensed matter phenomena \cite{QSim} and have recently been used to realize non-equilibrium transport processes, such as the conductance in mesoscopic channels \cite{Brantut2012} and its quantization \cite{Krinner2014}, thermoelectric transport \cite{Hazlett2013}, and the control of mobility in driven systems \cite{Salger2013}. Here, we study the transport of an interacting quantum gas through a multimode tunneling junction. The coherent interaction between the particles leads to a density dependent tunnel coupling, which gives rise to NDC. Our results give insight in the non-equilibrium dynamics at a tunneling junction and pave the road for the implementation of more complex atomtronic circuits \cite{Zoller2004,Eckel2014}.

The miminum instance of our experiment is depicted in Fig. 1A. Two particle reservoirs with chemical potential $\mu_1$ and $\mu_2$ are separated by a tunneling barrier and coherently coupled with strength $J$. Each subsystem contains many particles and provides many spatial modes. Preparing an initial non-equilibrium condition, $\Delta \mu=\mu_1-\mu_2\neq0$, we measure the ensuing transport through the tunneling barrier. Two physical mechanisms are important for the tunneling transport: (i) the interaction energy between the particles leads to a density dependent tunnel coupling and (ii) collisions between the particles provide an intrinsic source of decoherence.

We realize this model with a Bose-Einstein condensate of $^{87}$Rb atoms in a one-dimensional optical lattice \cite{Supplement}. Here, each lattice site is a two-dimensional Bose-Einstein condensate \cite{Hadzibabic2011} containing about 700 atoms. Employing a scanning electron microscopy technique \cite{Gericke2008,Wuertz2009}, we initially remove atoms from one site in a deep lattice where tunneling is absent (Fig. 1B), thus creating an imbalance in chemical potential $\Delta \mu$. Lowering the lattice height to different final values, we induce the transport in the array and observe the ensuing refilling dynamics. Fig. 1C shows the microscopic level structure of the tunneling junction. The chemical potential of the 2D condensate in a full site is much larger than the level spacing in the radial direction and many spatial modes contribute to the transport. In order to obey energy conservation, the particles can only tunnel into radially excited states of the central site. This goes along with a projection of the two radial wave functions, corresponding to an additional Franck-Condon factor in the tunneling matrix element \cite{Supplement}. The resulting effective tunnel coupling $J_{\mathrm{eff}}(\Delta \mu)$ depends on the difference in chemical potential between both reservoirs. With increasing $\Delta \mu$, the Franck-Condon factor and the effective tunnel coupling get smaller. After variable evolution times we probe the atom number and the transverse density distribution in the central site employing the same electron microscopy technique.

Fig. 2A shows the experimental results for different tunnel couplings $J$. In all cases the population in the central site smoothly recovers its equilibrium value. The absence of oscillations on top of the refilling curves is a first indicator for the presence of collisional decoherence in the central site, since purely coherent tunneling dynamics would lead to oscillations of the population, comparable to the Josephson oscillations between two coupled modes \cite{Albiez2005}. The density dependent tunneling matrix element directly affects the refilling dynamics. For lower values of $J$ the refilling starts visibly with a lower rate and gets faster over time, leading to an overall `s-shaped' form. This behavior can now be converted into a current-voltage relation, following Refs.\,\cite{Brantut2012,Krinner2014}. In Fig. 2B we plot the tunneling current defined as the derivative of the refilling dynamics against the difference in chemical potential $\Delta\mu$, representing the applied voltage. The `s-shape' translates into a non-monotonic dependence. For small $\Delta\mu$, the current increases linearly (see solid line in Fig. 2B), which is characteristic for an ohmic conductivity. It then bends over and the current is strongly suppressed for large differences in the chemical potential. These characteristics are known as NDC \cite{Esaki1958,ET1970,Pop2005,Britnell2013,Perrin2014}. It is technically exploited, e.g. in Gunn diodes for the generation of microwave radiation. In a tunnel diode as in most other devices, NDC occurs due to the interplay between the Fermi energy of the carriers on one side of the junction and the density of available hole states on the other side. The latter can be shifted out of resonance or can be depleted by an applied voltage. In superlattice structures, NDC is the result from a competition between localization length and collision rate. Here however, the microscopic origin of NDC is different: there is neither a detuning nor a change of localization. Instead, the NDC has its origin in a tunnel coupling which depends on the applied voltage.

The interaction between the atoms also affects the thermodynamic properties of the transport process. At the beginning of the transport the density distribution is dominated by thermal atoms, as the condition for condensation is barely reached. Upon tunneling, the interaction energy (given by the chemical potential $\mu_0$ of a full site) is converted into kinetic and potential energy, being thermalized after a few collisions. As the collision rate is in the order of a few hundred Hertz \cite{Supplement} thermal equilibrium is rapidly established. The depletion of the condensate in the full site is always balanced by the adjacent reservoirs sites and all atoms which tunnel into the central site have the same energy $\mu_0$. We can therefore apply the equipartition theorem to the 2D harmonic oscillator potential in the central site and find $\mu_0=2k_{\mathrm{B}} T$ for the temperature in the central site at the very first part of the dynamics. The prediction of $34(4)\,$nK is in excellent agreement with our experimental finding $36(2)\,$nK, extracted from a Gaussian fit to the radial density distribution (Fig. 3). This is an extreme manifestation of the Joule Thompson effect in the quantum regime \cite{Schmidutz2014}, where all interaction energy is converted into thermal energy. During the refilling process the width of the density distribution gets smaller, eventually reaching the equilibrium state.

The role of decoherence in transport processes has been extensively discussed in the literature \cite{Caldeira1981,Imry1999}. Its implication for our experiment can be understood from a microscopic description of the tunneling transport. To this purpose we use a master equation approach where the central site is subject to decoherence \cite{Supplement}. We fit our data with this model, leaving the decoherence rate $\Gamma_{\mathrm{dec}}$ as a free parameter. The results are shown as solid lines in Fig. 2A. The model describes our data very well for all tunnel couplings. We find a decoherence rate of about $600\,$Hz independent of the tunnel coupling $J$. Its magnitude is in accordance with the estimated thermal collision rate \cite{Supplement}. We therefore identify collisions as the dominant source of decoherence. As the collisions happen much faster than the transport process itself, thermal equilibrium is always established and we measure indeed a current as if the system was in a steady state at a given time. The decoherence rate is also larger than the tunnel coupling ($J/\hbar<400\,$Hz) and the transport is dominantly incoherent.

The importance of decoherence in our study manifests itself also in the appearance of the quantum Zeno effect \cite{Itano1990,Barontini2013}. It has been predicted that the effective hopping rate in such kind of system scales as $\tau_{\mathrm{eff}}^{-1}\propto J^2/\Gamma_{\mathrm{dec}}$ \cite{Kofman2000,Trimborn2011,Barmettler2011}. Analyzing the refilling time in dependence of the tunnel coupling for constant $\Gamma_{\mathrm{dec}}$ (Fig. 4A), we readily recover the quadratic power law: $\tau_{\mathrm{eff}}^{-1}\propto J^{\alpha}$ with $\alpha=1.9(1)$. In order to vary the decoherence rate $\Gamma_{\mathrm{dec}}$ we change the overall atom number, keeping the lattice depth constant ($J/\hbar=100\,$Hz). For $\Gamma_{\mathrm{dec}}>J/\hbar$, we find that $\tau_{\mathrm{eff}}\propto \Gamma_{\mathrm{dec}}$ (Fig. 4B). Both results together verify the predicted dependency: $\tau_{\mathrm{eff}}^{-1}\propto J^2/\Gamma_{\mathrm{dec}}$. Note that in the case of a 1D Josephson junction, the opposite behavior is expected \cite{Supplement}.

The observation of NDC in the transport of ultracold quantum gases bears great potential for future implementations of interaction controlled atomic circuits \cite{Eckel2014,Gutman2012,Ivanov2013}. It can also serve as an alternative mechanism for the observation of bistability and related non-equilibrium phenomena. Provided the two reservoirs have different spatial geometries, the transport process might become even asymmetric as a function of $\Delta \mu$, thus providing a new strategy to build a diode for neutral matter.


\bibliography{science}

\begin{thebibliography}{10}
\bibitem{Supplement}
Information on materials and methods is available at the Science Web site.
\end{thebibliography}

\bibliographystyle{Science}


\begin{scilastnote}
\item [\bf{Acknowledgments:}] We are grateful for useful discussion with James Anglin, Ennio Arimondo, Georgios Kordas, Artur Widera, and Dirk Witthaut. We acknowledge financial support by the DFG within the SFB/TR 49 and WI 3426/7. R. L. is supported by the MAINZ graduate school.
\end{scilastnote}


\clearpage

\begin{figure}
\begin{center}
\includegraphics[scale=1.0]{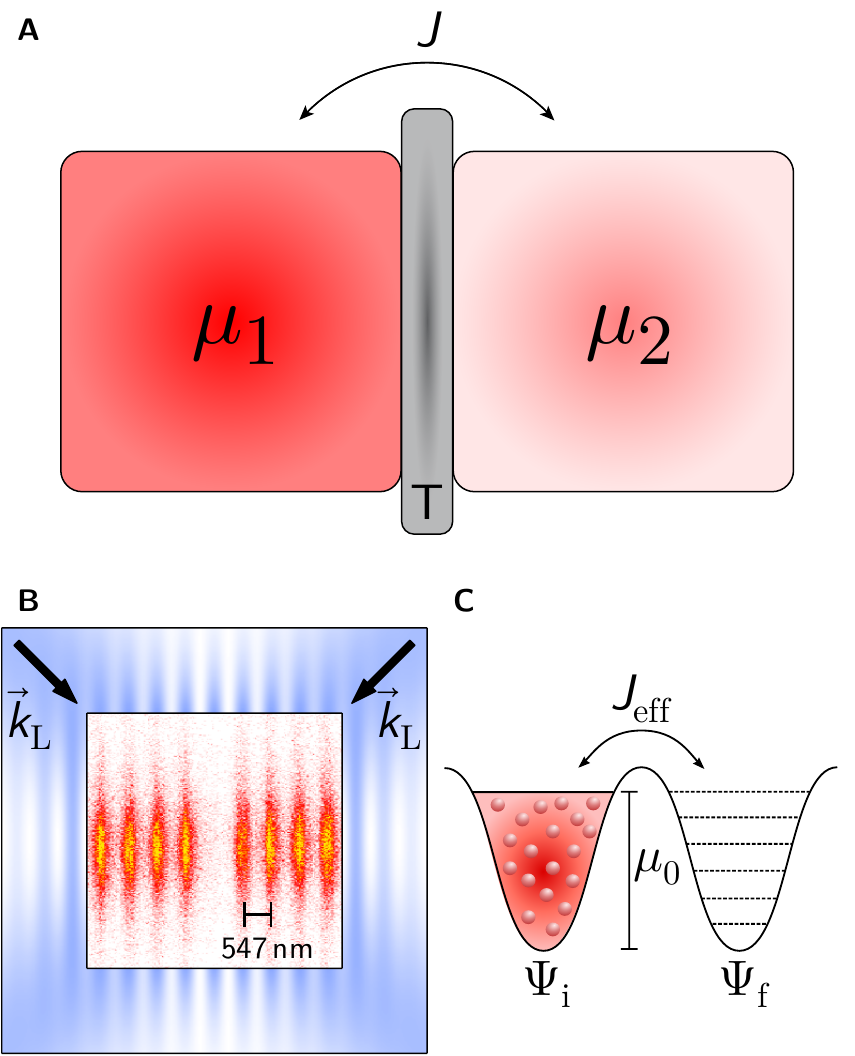}
\end{center}
\end{figure}

\paragraph*{\textsf{Fig. 1.}}
\textsf{(\textbf{\textsf{A}}) Two reservoirs can exchange particles with coupling strength \textit{J} through a tunneling barrier. The particles in both sides can occupy many spatial modes and have different chemical potentials. (\textbf{\textsf{B}}) Experimental setup: Two blue detuned laser-beams ($\vec{\textit{\textsf{k}}}_{\mathrm{L}}$) create a one-dimensional optical lattice in which we load a Bose-Einstein condensate. Removing atoms from the central site of this system leads to an out-of-equilibrium state as implementation of (\textbf{\textsf{A}}) (see text). (\textbf{\textsf{C}}) Microscopic view of the energy level structure. Particles of a full site with chemical potential $\mu_0$ are resonant with higher radial states of the empty site in which they tunnel with a reduced effective tunneling rate \it{\textsf{J}}$_{\mathrm{eff}}$.}

\clearpage

\begin{figure}
\begin{center}
\includegraphics[scale=1.0]{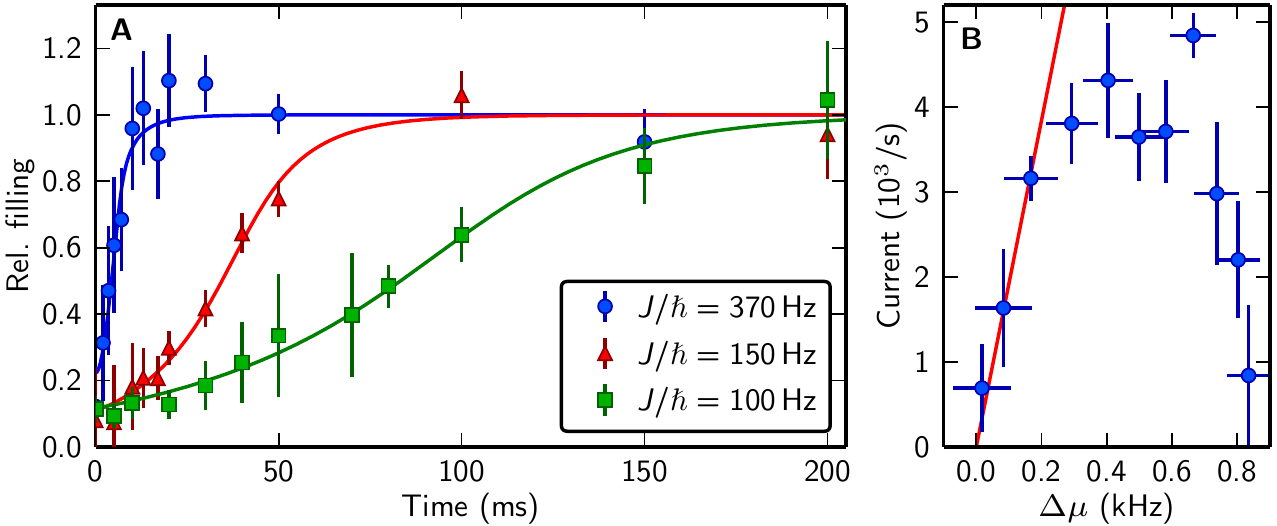}
\end{center}
\end{figure}

\paragraph*{\textsf{Fig. 2.}}
\textsf{(\textbf{\textsf{A}}) Refilling dynamics for three different tunnel couplings. The solid lines are the prediction of an effective theoretical model 
\cite{Supplement}. (\textbf{\textsf{B}}) Negative differential conductivity: after a linear initial rise (red line), the current drops for increasing imbalance in chemical potential.}

\clearpage

\begin{figure}
\begin{center}
\includegraphics[scale=1.0]{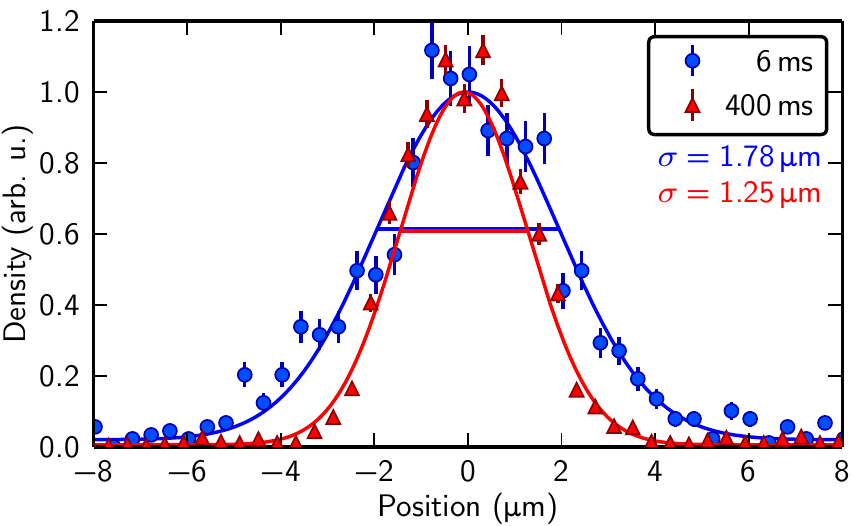}
\end{center}
\end{figure}

\paragraph*{\textsf{Fig. 3}}
\textsf{Normalized radial density distribution for the refilling dynamics at \textit{J}/$\hbar\;$=$\;$230$\,$Hz. After 6$\,$ms (blue dots; 220 atoms) the distribution is governed by thermal energy and we find a temperature of \textit{T}$\;$=$\;$36(2)$\,$nK by fitting a Gaussian function. The equilibrium state (red triangles; 600 atoms), which is reached after 400$\,$ms, is dominated by interaction energy and its widths gets smaller.}

\clearpage

\begin{figure}
\begin{center}
\includegraphics[scale=1.0]{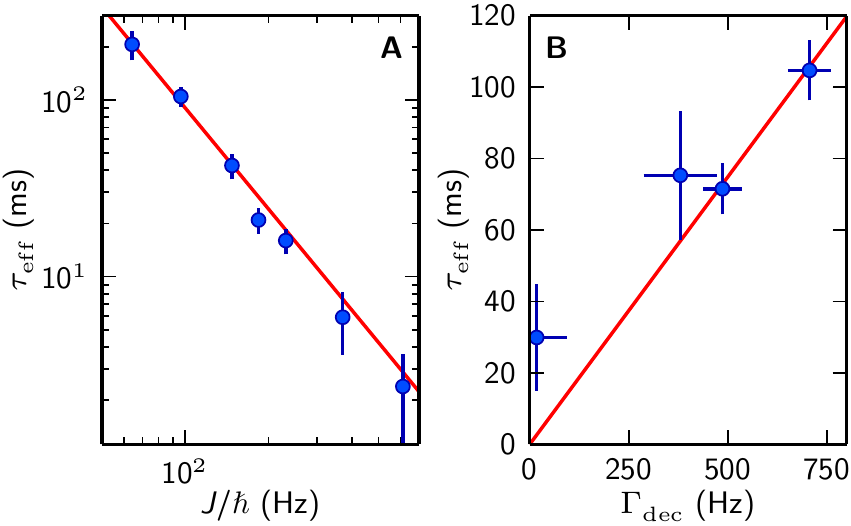}
\end{center}
\end{figure}

\paragraph*{\textsf{Fig. 4}}
\textsf{(\textbf{\textsf{A}}) Refilling time versus tunnel coupling for constant decoherence rate. We fit a power law with an exponent of $\alpha\;$=$\;$1.9(1). (\textbf{\textsf{B}}) Refilling time versus decoherence rate: manifestation of the quantum Zeno effect. $\tau_{\mathrm{eff}}$ scales linear with $\Gamma_{\mathrm{dec}}$ (red line).}

\clearpage

\section*{Supplementary Materials}
\begin{itemize}
\item[] Materials and Methods
\item[] Figures S1-S2
\item[] References (34-40)
\end{itemize}

\paragraph*{Experimental setup}
We create a Bose-Einstein condensate (BEC) of $^{87}$Rb with about $45\times10^3$ atoms in a single beam dipole trap realized by a CO$_2$-laser \cite{Gericke2007}. We then load the BEC in a one-dimensional optical lattice created by two blue detuned laser beams ($\lambda=774\,$nm) crossed at $\theta = 90^\circ$. The resulting trap-frequencies in a lattice site are $\nu_r=165\,$Hz and $\nu_{z}=40-100\,$kHz and each site contains about 700 atoms. 

\paragraph*{Franck-Condon factor}
The atoms with chemical potential $\mu_0$ can only tunnel resonantly into higher orbital states. We determine the effective tunnel coupling by calculating the overlap between the initial wave function at the full site and that of a 2D harmonic oscillator. Fig. S1 shows the measured wave function in a full site and the first few harmonic oscillator states in the radial direction. Since the atoms in the full site have no angular momentum with respect to the axial symmetry axis, only states with $m=0$ are accessible and therefore restricts the quantum number $n$ to be even. We calculate the additional Franck-Condon factor as 
\begin{equation}
\eta = \left<\psi_{\mathrm{exp}}|\psi_{n}\right>
\end{equation}
where $n$ is chosen to be the nearest available orbital state that leads to resonant tunneling between the sites. The tunnel coupling to an empty site is therefore reduced and given by $\eta\cdot J$. 

\paragraph*{Theoretical model}
To describe our experimental data we start with a master equation in Lindblad form \cite{Breuer2002,Kordas2012,Kordas2013}
\begin{equation}
i\hbar\partial_t\hat{\rho}=\left[\hat{H}_{\mathrm{eff}},\hat{\rho}\right]+i\hbar\hat{\mathcal{L}}\hat{\rho}\, ,
\end{equation}
with $\hat{\rho}$ the density matrix and $\hat{H}_{\mathrm{eff}}$ an effective single-particle Hamiltonian where we mimic the many-particle behavior of the system with help of the effective tunnel rate $J_{\rm eff}(\Delta\mu)$, that couples the central site with its neighbors. The coupling between all other sites is given by the unchanged tunneling matrix element since they have the same filling. To model the effective tunnel coupling we assume it to be linear in the difference of atom number with the minimal value of $\eta\cdot J$ when one site is empty. In the parameter regime of the experiment the chemical potential increases linearly with the atom number, which allows us to calculate $J_{\mathrm{eff}}\left(\Delta\mu\right)$ as follows: 
\begin{equation}
J_{\mathrm{eff}}\left(\Delta\mu\right) = J - \left(1-\eta\right)J \cdot \Delta \mu/\mu_0 \times 1.7 \,, 
\end{equation}
where $\Delta\mu$ is the difference in chemical potential. The factor $1.7$ arises from the conversion of atom number into chemical potential. Note that Eq. (3) is only applicable for $\Delta\mu/\mu_0 < 0.7$ as it is the case for our experiment.

We model the Liouville superoperator as follows: within the central site, many quantum states in the radial direction are accessible and the atoms can scatter between them. As the available many-body Hilbert space is huge and as the collision rate is much higher than the coherent coupling, we assume thermodynamic equilibrium. It has recently been shown \cite{Gilz2014}, that in such a situation, the collisional rate in a thermal ensemble of particles is equal to the decoherence rate. We therefore define the Liouville superoperator for our effective model as \cite{Witthaut2011}
\begin{equation}
\hat{\mathcal{L}}\hat{\rho}=\Gamma_{\mathrm{dec}}\left(2\hat{n}\hat{\rho} \hat{n}^{\dagger}-\hat{n}^2 \hat{\rho} - \hat{\rho} \hat{n}^2\right)
\end{equation}
with $\hat{n}$ the number operator acting solely on the central site. The good agreement between this model and the experiment shows that even though the global system is closed, the effective description as an open system works well and the intrinsic collision processes can be regarded as Markovian.

\paragraph*{Collision rate}
The collision rate for an atom which tunnels into the central site has two contributions. One is the thermal collision rate which can be calculated for a 2D-Bose gas close to condensation \cite{Guery-Odelin2004} and which ranges from $150\,$Hz to $200\,$Hz for the different values of $J$. A second effect comes from the accumulation of atoms in only one or two radially excited states. As the population in these states is much larger than one, the interaction energy couples the atoms efficiently to neighboring states. This opens additional scattering channels through which the atoms can leave the original state. We estimate a contribution of a few hundred Hertz for this process. As the critical density for Bose-Einstein-condensation within the central site is already reached for about 50 atoms, the population in excited states is always saturated and the condensate fraction increases during refilling. Only atoms in excited states contribute to collisional decoherence. We therefore assume in our model that the decoherence rate is constant.

\paragraph*{Refilling time}
We define the effective refilling time as the point where the filling exceeds $\sim 67$\% analogous to the definition of $\tau$ in an exponential function. For our data this can be extracted from the fitted simulations.

\paragraph*{Josephson junction}
It is instructive to compare our results to the 1D situation of a two-mode Josephson junction \cite{Albiez2005}. We have performed numerical simulations of a three-well Josephson junction in a full many-body setup described by Eqs. (1) and (3) but with uniform decoherence in all three wells, see \cite{Kordas2012,Kordas2013} for similar model computations. Fig. S2 shows examples of the refilling dynamics for different strengths of the decoherence. The oscillations between the wells are efficiently damped and a transport behavior is found similar to the one studied here experimentally. However, for increasing decoherence, the refilling becomes faster. This is a manifestation of the anti Zeno effect \cite{Kofman2000}. Even in the extreme case of self-trapping, the tunneling junction can be made conductive by implementing a sufficiently strong source of decoherence. Instead, in our experiment, we observe Zeno dynamics due to the 2D geometry of our setup, in which self-trapping is prohibited by the additional radial degree of freedom (cf. discussion above).

\clearpage

\begin{figure}
\begin{center}
\includegraphics[scale=1.00]{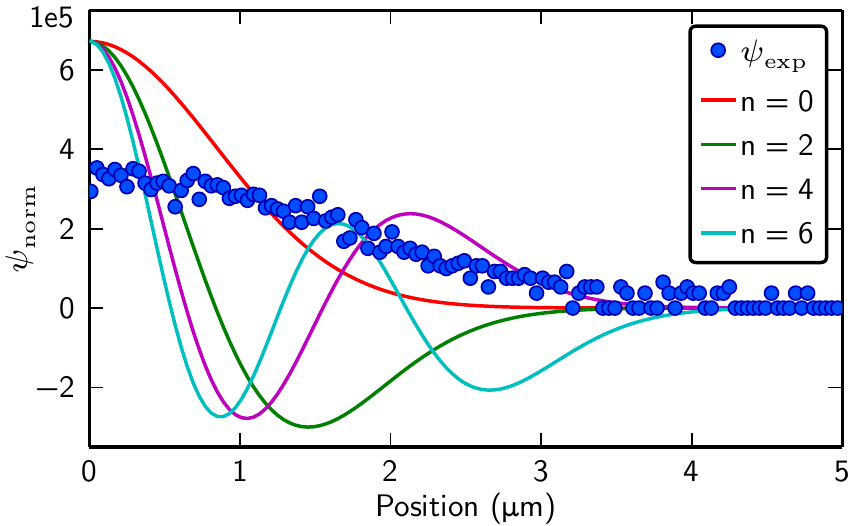}
\end{center}
\end{figure}

\paragraph*{\textsf{Fig. S1}}
\textsf{Radial wavefunctions to calculate the effective overlap. The dots show the experimental radial wavefunction extracted from the measured density distribution. The solid lines are the calculated wavefunctions of a 2D harmonic oscillator for m$\;$=$\;$0 in a polar basis.}

\clearpage

\begin{figure}
\begin{center}
\includegraphics[scale=1.00]{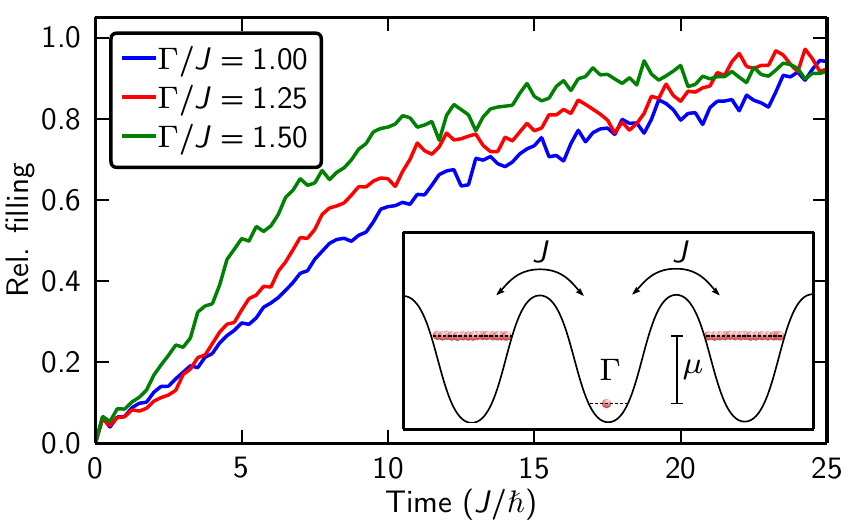}
\end{center}
\end{figure}

\paragraph*{\textsf{Fig. S2}}
\textsf{Many-body simulations of a three-well 1D Josephson junction for three different dephasing rates $\Gamma$ and 30 atoms.
With increasing interaction the tunneling process becomes off-resonant while the coupling itself is not changed. The refilling time gets shorter with increasing decoherence, indicating anti Zeno dynamics.}
\end{document}